\documentclass[11pt]{article}
\usepackage[utf8]{inputenc}
\usepackage{lipsum}
\usepackage{hyperref}
\usepackage{enumitem}
\usepackage{float}
\usepackage{graphicx}
\usepackage{caption} 
\usepackage{calc}
\usepackage{fancyhdr}
\usepackage{tcolorbox}
\usepackage{hyperref}
\usepackage{wrapfig}
\usepackage{graphicx} % Required for inserting images
\usepackage{amssymb}
\usepackage{amsmath}
\usepackage{multicol}
\usepackage{braket}
\usepackage{abstract}
\usepackage{multicol}
\usepackage{xcolor}
\usepackage{appendix}
\title{\normalsize\bf Constraints on BMS Transformations via Energy Conditions and \\
implications on black hole geometry}
\author{\normalsize {\sc Nihar Ranjan Ghosh\footnote{\tt g.nihar@iitg.ac.in}\ \ and Malay K. Nandy\footnote{\tt mknandy@iitg.ac.in}}\\
\normalsize \em Department of Physics, Indian Institute of Technology Guwahati\\
\normalsize \em Guwahati 781 039, India}
\date{March 6, 2026}

\begin{document}

\maketitle

\begin{abstract}

We investigate whether the formally infinite-dimensional supertranslation sector of the Bondi–Metzner–Sachs (BMS) group remains fully physically admissible once classical energy conditions are enforced. Working in a perturbative framework $g_{ab}\to g_{ab}+h_{ab}$, we first develop a general  toolkit by expanding the curvature tensors and the Ricci scalar in powers of the perturbation $h_{ab}$ and recast the strong, weak, null and dominant energy conditions (SEC, WEC, NEC and DEC, respectively) as explicit inequalities on $h_{ab}$ following from  the Raychaudhuri equation.  The formalism is general, but to obtain concrete constraints we specialize to the standard BMS form on a Schwarzschild background and parametrize $h_{ab}=\mathcal{L}_\eta g_{ab}  $ by a supertranslation function \(f(\theta,\phi)\). We find that the SEC and WEC impose nontrivial angular restrictions on \(f\) already at next-to-leading order (NLO) in the perturbation, whereas the NEC and DEC are preserved at linear order and acquire their first nontrivial contributions only at next-to-next-to-leading order (NNLO). Notably, the NNLO NEC reduces to a purely angular condition (independent of the radial coordinate), providing the strongest constraint on admissible supertranslations.  Thus, imposing energy
conditions substantially reduces the space of physically admissible supertranslations; the
allowed sector, although remains infinite-dimensional in principle, is substantially constrained
in practice.
\end{abstract}

\maketitle

\section{Introduction}

The Bondi-van der Burg-Metzner-Sachs (BMS) symmetry group \cite{bondi1962gravitational,sachs1962gravitational} represents the infinite dimensional symmetry group of asymptotically flat spacetimes, extending the finite-dimensional Poincar\'e group by an infinite number of generators known as {\em supertranslations}. These transformations can also be interpreted by angle-dependent translations near null infinity, and they significantly enrich the phase space structure of general relativity by distinguishing between spacetime configurations that are otherwise indistinguishable under Poincar\'e symmetry \cite{ghosh2025asymptoticgenerationkerrgeometry}.

This infinite dimensional symmetry group was the unexpected outcome of the work by Bondi, van der Burg, Metzner, and Sachs's search for the non-trivial symmetry group of asymptotically flat spacetimes. Contrary to the common expectations of the isometries of Minkowski space, these infinite number of generators was discovered, vastly enriching the symmetry structure near null infinity with its further and deep connections to the so called {\em infrared triangle} \cite{strominger2018lecturesinfraredstructuregravity}, whose three corners comprise asymptotic symmetries, soft theorems, and gravitational memory effects.

Similar development occurred in the context of asymptotically anti-de Sitter (AdS) spacetimes, notably through the work of Brown and Henneaux \cite{brown1986central}, who demonstrated that the asymptotic symmetry algebra of $\mathrm{AdS}_3$ gravity consists of two copies of the Virasoro algebra. While the focus of this work remains on asymptotically flat spacetimes, the structural similarities in both settings highlight the universality and physical relevance of asymptotic symmetry analysis \cite{CADONI1999165, hotta1998asymptoticisometrydimensionalantide, henneaux1985asymptotically, Comp_re_2016}.

\subsection{Gravitational Memory and the Infrared Triangle}
Within the infrared triangle, the {\em gravitational memory effect} offers a classical and observable manifestation of BMS symmetry. First discovered in the linearized regime by Zel'dovich and Polnarev \cite{Zeldovich:1974gvh} and subsequently extended to the nonlinear regime by Christodoulou and others \cite{Christodoulou, Braginsky:1985vlg, Braginsky:1987kwo, PhysRevD.44.R2945, PhysRevD.46.4304, PhysRevD.45.520, Favata:2010zu, Tolish:2014bka, Tolish:2014oda, Winicour:2014ska}, the memory effect describes the permanent displacement of freely falling test masses due to the propagation of gravitational waves. This permanent displacement is direct link to BMS supertranslations, as the memory effect encodes the  radiative history of the spacetime \cite{Strominger:2014pwa, strominger2018lecturesinfraredstructuregravity, PhysRevD.92.084057, Flanagan:2015pxa, Pasterski:2015tva}.

Moreover, recent work has enlarged the notion to include spin and center of mass memory effects \cite{Pasterski:2015zua, PhysRevD.98.064032, Compere:2016jwb}, as well as the electromagnetic analogues \cite{Bieri:2013hqa, susskind2019electromagneticmemory}. In the contexts of black holes, Donnay et al.~\cite{PhysRevLett.116.091101} introduced the notion of black hole memory, showing that transient radiation can produce permanent deformations of the near horizon geometry, resulting in a route between the memory effect to the emergence of soft hair on horizons. Studies of extremal black holes further reveal distinctive memory related features arising from horizon degeneracy \cite{PhysRevD.102.044041}.

\subsection{Soft Theorems and Asymptotic Symmetries}

The third component of the infrared triangle, the {\em soft theorems}, originates from early studies in quantum electrodynamics, notably through the work of Bloch and Nordsieck \cite{PhysRev.52.54}, Low \cite{Low:1954kd, Low:1958sn}, Gell-Mann and Goldberger \cite{Gell-Mann:1954wra}, and Yennie et al.\ \cite{Yennie:1961ad}. The gravitational realisation of  the soft theorem was developed by Weinberg \cite{Weinberg:1965nx}, who showed that the emission of low-energy gravitons in scattering processes is governed by universal behavior independent of the specific details of the interacting fields.

Crucially, these universal behaviors are now understood to correspond to Ward identities
associated with asymptotic symmetries, providing a direct correspondence between soft theorems and BMS charges \cite{Pasterski:2015tva}. In this light, soft theorems offer a quantum field theoretic realization of the infinite dimensional symmetry structure revealed by BMS analysis and serves as the fundamental idea behind {\em soft hair} in black-hole physics.

\subsection{Soft Hair and the Information Paradox}

The soft-hair proposal has drawn significant interest because of its potential applications in the black hole information loss paradox \cite{PhysRevD.14.2460}. Hawking, Perry, and Strominger \cite{PhysRevLett.116.231301} argued that the infinite number of conserved BMS charges corresponds to infinitely many zero-energy excitations, the soft hair, associated with the black-hole horizon. These soft degrees of freedom could, in principle, store information about black hole microstates and hence provide a channel for information retention during Hawking evaporation \cite{Hawking:1975vcx, Hawking:2016sgy, haco2018black, PhysRevD.96.084032, Chu_2018, PhysRevD.108.044034}.

This proposal significantly modifies the classical no-hair theorem \cite{Israel:1967za, Israel.164.1776, PhysRevLett.26.331}, which asserts that black holes are characterized by only a finite set of global charges. The inclusion of soft hair reveals an infinite-dimensional structure not accounted for in the traditional framework \cite{strominger2017blackholeinformationrevisited}, and has important consequences for black hole entropy and unitarity in quantum gravity \cite{haco2019kerrnewmanblackholeentropy, PhysRevD.103.126020}.

An alternative but complementary viewpoint treats soft hair as edge modes, which are boundary-localized degrees of freedom that arise from large gauge transformations and acts nontrivially at spacetime boundaries. These edge modes are crucial for maintaining gauge invariance on subregions and have been extensively studied in gauge theories and gravity \cite{Donnelly:2014fua, Donnelly:2015hxa, Harlow:2015lma, Harlow:2016vwg, Maldacena:2016upp}, suggesting that soft hair may be a generic feature of gauge theories with boundaries.

\subsection{Constraints on the BMS Transformation}
In light of these recent and vast developments, it is legitimate to ask whether the infinite dimensional BMS symmetry group is indeed infinitely vast or not. Apart from preserving the asymptotic behaviour, for example flatness, and causality of the spacetime, it is imprtant to ask whether there exist other conditions  applicable to the infinite number of generators that can  reduce the number of generators or impose constrains on them. Some of such obvious and mandatory constraints can be implemented via the energy conditions.

It is well-established that the energy conditions have been employed across diverse scenarios to obtain a broad, model independent results. For instance, the Hawking-Penrose singularity theorems utilize the weak energy condition (WEC) and strong  energy condition (SEC), while the proof of the black hole area theorem requires the null energy condition (NEC). Recently, these classical energy conditions of general relativity have also been applied to various cosmological problems, including analyses of phantom field potentials \cite{santos2005energy} and expansion history of the universe \cite{Santos_2007, Sen_2008}, as well as in constraining different $f(R)$ models \cite{Santos:2007bs}. Importantly, these energy conditions were formulated in the framework of general relativity. Nevertheless, these conditions can be mapped to a geometry obtained through the action of large gauge transformation on a geometry which is a solution of the  Einstein's field equation, such as the  Schwarzschild solution.

Thus, if one starts with an unperturbed metric $g_{ab}$, which through some diffeomorphic vector field $\eta$ is transformed to the new metric $\bar{g}_{ab}$ via $ds^2=\bar{g}_{ab}dx^adx^b=(g_{ab}+\mathcal{L}_\eta g_{ab})dx^adx^b$, one can map these conditions into  SEC, WEC, NEC, and dominant energy condition (DEC) with respect to the new metric $\bar{g}_{ab}$. Fulfilment of these four energy conditions will naturally impose constraints on the diffeomorphic vector field $\eta$ and will certainly reduce the size of the symmetry group. Furthermore, meeting these four conditions is necessary for gravity to be attractive, energy density to be positive and the transformations to be physically acceptable. Thus it is of outmost importance to impose the energy conditions on the metric field $\bar{g}_{ab}$, originating from  BMS transformations on $g_{ab}$, generated by a vector field $\eta$, which in turn reduces the size of the symmetry group.

Furthermore, for the sake of viability of the procedure, it is necessary to derive the SEC and NEC for a general metric expressed in BMS form  employing the Raychaudhuri equation, with the additional requirement that gravity be attractive in the $\bar{g}_{ab}$ metric.

In this work we pursue the above program for a general metric $g_{ab}$  expressed in standard BMS coordinates $(u,r,\theta,\phi)$. We construct the Ricci tensor $ \bar R_{ab}$ and Ricci scalar $\bar R$ of the perturbed metric $\bar g_{ab}$ and impose the energy conditions to obtain the necessary constraints on the generator $\eta^{a}$.
Furthermore, we express $\bar R_{ab}$, $\bar g^{ab}$, and $\bar R$ in terms of the original metric $g_{ab}$ and the perturbation $h_{ab}=\mathcal{L}_{\eta}g_{ab}$ by expanding them in leading order of $h_{ab}$ and by employing the energy conditions in the original and transformed descriptions.
This analysis can be generalised to any finite order $\mathcal{O}(h_{ab}^n)$.

\subsection{Raychaudhuri Equation and Energy Conditions}{\label{sec-raychaudhuri}}
As a starting point of our analysis to extend the SEC and NEC for arbitrary spacetime metric, we shall energy conditions followed by the timelike Raychaudhuri equation, with time like four vector $t^a$ as 
\begin{equation}
    \label{ray0}
    \frac{d\Theta}{d\tau}=-\frac{1}{3}\Theta^2-\sigma_{ab}\sigma^{ab}+\omega_{ab}\omega^{ab}-R_{ab}t^at^b
\end{equation}
where $\Theta$, $\sigma_{ab}$ and $\omega_{ab}$ are respectively the expansion parameter, shear, and rotation parameter associated with the congruence defined by the timelike vector field $t^a$.
Since shear is a spatial tensor, we must have $\sigma_{ab}\sigma^{ab}\geq 0$, and one can always choose a hypersurface orthogonal congruences such that $\omega_{ab}\omega^{ab}=0$. As $\Theta^2\geq 0$, the requirement for gravity to be always  attractive   ($\frac{d\Theta}{d\tau} \leq 0$) suggests that $R_{ab}t^at^b\geq 0$  for any timelike congruence.  This last inequality is known as the strong energy condition.

Importantly, being a geometric identity, the Raychaudhuri equation is independent of any particular theory, and the Ricci tensor $R_{ab}$ gets related to the energy momentum tensor $T_{ab}$ only when the Einstein's field equations are used. Therefore, the SEC can also be written as 
\begin{equation}
    R_{ab}t^at^b=\left(T_{ab}-\frac{1}{2}T g_{ab}\right)t^at^b\geq 0
\end{equation}
Thus, by expressing the transformed  Ricci tensor $\bar{R}_{ab} $ and the corresponding Ricci scalar $\bar{R}$ defined by the transformed metric $\bar{g}_{ab} $ in terms of $R_{ab}$ and $R$ defined in the old metric $g_{ab}$, we can identify $\bar{R}_{ab}-\frac{1}{2}\bar{R}\bar{g}_{ab} $ with an effective energy momentum tensor $\Theta_{ab}$ upon which the energy conditions can be implemented. 

Moreover, the Raychaudhuri equation for a null like vector field $n^a$ has the same structure as equation \ref{ray0} but with
a factor $1/2$ rather than $1/3$ in the first term on right hand side, and with $-R_{ab}n^an^b$ as the last term. Thus, geodesic focusing  
suggests that $R_{ab}n^an^b\geq 0$, known as the null energy condition (NEC). Just like for SEC, the same concept of effective energy momentum tensor can be applied here to generalize the condition for any theory of gravity, as was done for $f(R)$ theories in \cite{Santos:2007bs}.

The remainder of this paper is organized as follows. In Sec.~\ref{toolkit} we develop a general  toolkit for metric perturbations, including the second-order expansion of the curvature tensors and Ricci scalar, together with a systematic formulation of the energy conditions applicable to arbitrary perturbations.
In Sec.~\ref{sec-bms energy} we specialize this formalism to BMS transformations of the Schwarzschild background. We derive the constraints on the BMS supertranslation function that follow from imposing the energy conditions on the transformed geometry and analyze the physical implications of the SEC, WEC, DEC, and NEC in this setting.
Finally, in Sec.~\ref{disc} we conclude the paper with a discussion.

\section{General toolkit for metric perturbation}{\label{toolkit}}
In this section we shall develop a general toolkit for metric perturbation given by $\bar{g}_{ab}=g_{ab}+h_{ab}$ in order to develop constraints on the perturbation $h_{ab}$ imposing different energy conditions.

\subsection{Second order expansion of curvature tensors}{\label{sec-series expansion}}
As mentioned earlier, to express the new Ricci tensor and Ricci scalar in terms of the old ones (defined in terms of the unperturbed metric) we shall expand different curvature tensors up to linear order in the perturbation, $\mathcal{O}(h_{ab})$. Below we give a brief discussion about the expansion of different curvature tensor up to second order $\mathcal{O}(h_{ab}^2)$.

Consider the perturbed metric $\bar{g}_{ab}$ to be obtained in terms of the unperturbed metric $g_{ab}$, with  perturbation $h_{ab}$, 
\begin{equation}
    \label{metric perturbation}
    \bar{g}_{ab}=g_{ab}+\tau h_{ab}~,
\end{equation}
where $\tau$ is a book keeping  parameter, inserted to keep track of the order of expansion, which will be set to unity at the end of calculation. The inverse of the metric ${\bar g}^{ab}$ takes the form
\begin{equation}
    \label{inverse metric}
    {\bar g}^{ab}=g^{ab}-\tau h^{ab}+\tau^2 h^{ac} h^b_c~+\mathcal{O}(\tau^3).
\end{equation}
Raising and lowering of indices is therefore implemented with respect to $g_{ab}$, and the trace of the metric perturbation is given by $h=g^{ab}h_{ab}$. Similarly, the second order expansion of the Christoffel symbol is given by 
\begin{equation}
    \label{christoffel}
    \bar{\Gamma}^a_{bc}=\Gamma^a_{bc}+\tau (\Gamma^a_{bc})_L-\tau^2 h^a_e (\Gamma^e_{bc})_L+\mathcal{O}(\tau^3)~,
\end{equation}
with the linearized Christoffel symbols defined as
\begin{equation}
    \label{lchristoffel}
    (\Gamma^a_{bc})_L=\frac{1}{2}g^{ae}\left[ \nabla_b h_{ec}+\nabla_c h_{be}-\nabla_e h_{bc}   \right]~,
\end{equation}
where the $\nabla$'s are defined with respect to the background metric $g_{ab}$ and $\Gamma^a_{bc}$ are the Christoffel symbols compatible with the background metric, i.e, $\nabla_b g_{bc}=0$.

The Riemann tensor in the transformed metric $\bar{g}_{ab}$ is given by
\begin{equation}
    \label{riemann}
    \bar{R}^a_{~~bcd}=R^a_{~~bcd}+\tau(R^a_{~~bcd})_L-\tau^2h^a_e(R^e_{~~bcd})_L-\tau^2g^{ai}g_{ej}\left[(\Gamma^j_{ci})_L(\Gamma^e_{db})_L-(\Gamma^j_{di})_L(\Gamma^e_{cb})_L  \right]+\mathcal{O}(\tau^3)
\end{equation}
 with 
 \begin{equation}
     (R^a_{~~bcd})_L=\frac{1}{2}\left[\nabla_c\nabla_dh^a_b+\nabla_c\nabla_bh^a_d-\nabla_c\nabla^ah_{db}-\nabla_d\nabla_ch^a_b-\nabla_d\nabla_bh^a_c+\nabla_d\nabla^ah_{bc}   \right]
 \end{equation}
Thus, the second order expansion for the Ricci tensor is 
\begin{equation}
    \label{ricci tensor}
    \bar{R}_{bd}=R_{bd}+\tau (R_{bd})_L-\tau^2h^a_e(R^e_{~~bad})_L-\tau^2g^{ai}g_{ej}\left[(\Gamma^j_{ai})_L(\Gamma^e_{db})_L-(\Gamma^j_{di})_L(\Gamma^e_{ab})_L  \right]+\mathcal{O}(\tau^3)
\end{equation}
with 
\begin{equation}
    (R_{bd})_L=\frac{1}{2}\left[ \nabla_e\nabla_bh^e_d+\nabla_e\nabla_dh^e_b-\nabla_e\nabla^eh_{bd}-\nabla_b\nabla_d h  \right]~,
\end{equation}
giving the transformed Ricci scalar,
\begin{equation}
    \label{ricci}
    \begin{split}
        \bar{R}=R+\tau R_L&+\tau^2\Big[ R^{ab}h_{ac}h^c_b-h^{ab}(R_{ab})_L-g^{ab}h^c_d(R^d_{~~acb})_L\\
        &-g^{ab}g^{cd}g_{ei}\Big\{(\Gamma^i_{cd})_L(\Gamma^e_{ba})_L -(\Gamma^i_{bd})_L(\Gamma^e_{ca})_L \Big\}    \Big]+\mathcal{O}(\tau^3)
    \end{split}
\end{equation}
where 
\begin{equation}
    R_L=g^{ab}(R_{ab})_L-R^{ab}h_{ab}~~.
\end{equation}
Thus, from equations \ref{ricci tensor} and \ref{ricci} the transformed Einstein   tensor $\bar{G}_{ab}(\bar{g}_{ab})$ is expressed as 
\begin{equation}
    \label{einstein}
    \begin{split}
        \bar{R}_{ab}-\frac{1}{2}\bar{R}\bar{g}_{ab}=
    \left(R_{ab}-\frac{1}{2}R g_{ab}\right)&+\tau\frac{1}{2}\Big[ \nabla_c\nabla_bh^c_a+\nabla_c\nabla_a h^c_b-\square h_{ab}-\nabla_b\nabla_ah-R h_{ab}\\
    &-\left(\nabla_c\nabla_dh^{cd}-\square h-R_{cd}h^{cd}  \right)g_{ab} \Big]+\mathcal{O}(\tau^2)
    \end{split}
\end{equation}
  up to order $\mathcal{O}(h)$.

Now, if one starts with an action $\bar{S}=\frac{1}{2}\int\sqrt{-\bar{g}}~ d^4x \bar{R}+\bar{S}_M$, one obtains the Einstein equation as $\bar{R}_{ab}-\frac{1}{2}\bar{R} \bar{g}_{ab}=\Theta_{ab}  $, where  $\Theta_{ab}$ is the energy momentum tensor of the matter field. Consequently, the right hand side of equation \ref{einstein}  can be interpreted as the effective energy momentum tensor $\Theta_{ab}$ in the transformed frame.

\subsection{Energy Conditions} \label{sec-energy}

From a physical standpoint, any transformation acting on a spacetime metric cannot be regarded as innocuous unless it preserves the fundamental viability conditions that define reasonable classical matter configurations. In general relativity, these viability criteria are encoded in the classical energy conditions, which serve as consistency requirements linking geometry to physically acceptable stress-energy content.

The classical energy conditions can be imposed directly on the transformed metric $\bar g_{ab}$. Requiring these conditions to remain satisfied after the action of a general transformation  $\bar g_{ab}=g_{ab}+h_{ab}$ yields nontrivial restrictions on $h_{ab}$,  thereby constraining the physically admissible sector of the transformation.

Therefore, using equation \ref{ricci tensor}, the SEC $\bar{R}_{ab}t^at^b\geq0 $ gives
 \begin{equation}
     \label{SEC}
     \Big[R_{ab}+\tau\frac{1}{2}\left\{2\nabla_c\nabla_ah^c_b-\square h_{ab}-\nabla_a\nabla_bh  \right\}     \Big]t^at^b+\mathcal{O}(\tau^2)\geq 0~.
 \end{equation}
Moreover, using equation \ref{einstein} the WEC $\Theta_{ab}t^at^b=(\bar{R}_{ab}-\bar{R}\bar{g}_{ab}/2)t^at^b\geq0$ gives
\begin{equation}
    \label{WEC}
    \begin{split}
       \Bigg[ \left(R_{ab}-\frac{1}{2}R g_{ab}   \right)&+\tau\frac{1}{2}\Big\{2\nabla_c\nabla_ah^c_b-\square h_{ab}-\nabla_a\nabla_bh -Rh_{ab}\\
       &-\left( \nabla_c\nabla_dh^{cd}-\square h-R_{cd}h^{cd} \right)g_{ab}   \Big\}\Bigg]t^at^b+\mathcal{O}(\tau^2)\geq0
    \end{split}
\end{equation}

Henceforth, we shall assume $t^a$ to be defined with respect to $\bar{g}_{ab}$ so that  $\bar{g}_{ab}t^at^b=-1$.

Similarly, the DEC $\Theta_{ab}\Theta^b_c t^at^c\leq 0 $ gives
\begin{equation}
    \label{DEC}
    \begin{split}
        &\left[R^c_aR_{cb}-R R_{ab}+\frac{1}{4}R^2 g_{ab}  \right]t^at^b+\tau \Bigg [ R^c_b \Big\{\nabla_d\nabla_ah^d_c-\nabla_d\nabla_ch^d_a-\square h_{ac} -\nabla_a\nabla_c h-R h_{ac}  \Big\}\\
        &-\Big\{\nabla_c\nabla_ah_b^c-\frac{1}{2}\square h_{ab}-\frac{1}{2}\nabla_a\nabla_b h\Big\}-\left(\nabla_c\nabla_d h^{cd}-\square h-R_{cd}h^{cd}  \right)\left(R_{ab}-\frac{1}{2}R g_{ab}\right)          \Bigg]t^at^b+\mathcal{O}(\tau^2)\leq 0
    \end{split}
\end{equation}
Lastly, the NEC $(\Theta_{ab}-\Theta g_{ab}/2)n^an^b\geq 0$, for any null four velocity $\bar{g}_{ab}n^an^b=0$ gives
\begin{equation}
    \label{NEC}
    \Big[R_{ab}+\tau\frac{1}{2}\left\{  2\nabla_c\nabla_ah^c_b-\square h_{ab}-\nabla_a\nabla_bh\right\}\Big]n^an^b+\mathcal{O}(\tau^2)\geq 0~,
\end{equation}

The four conditions \ref{SEC}, \ref{WEC}, \ref{DEC}, and \ref{NEC} provide the necessary constraints that a metric perturbation must satisfy, up to order $\mathcal{O}(h)$, in order to be physically admissible. Satisfaction of Eq.~\ref{SEC} ensures the attractive nature of gravity, while Eq.~\ref{WEC}, following Hawking and Ellis \cite{Hawking:1973uf}, guarantees that the energy density measured by any timelike observer is non-negative. Equation~\ref{DEC} further requires non-negative energy density together with a causal (non-spacelike) energy flux vector. The null energy condition, Eq.~\ref{NEC}, is associated with the boundedness from below of the corresponding Hamiltonian \cite{RANJANGHOSH2025117234}.

As will be demonstrated in the following section, for the Schwarzschild background subject to the class of BMS transformations considered here, the left-hand sides of Eqs.~\ref{DEC} and \ref{NEC} vanish identically at order $\mathcal{O}(h)$. The first nonvanishing contributions arise at order $\mathcal{O}(h^{2})$. For completeness, we therefore present the DEC and NEC constraints consistently up to second order in the perturbation.

Therefore, the DEC at order $\mathcal{O}(h^2)$ gives
\begin{equation}
    \label{2nd order DEC}
    \left[ Q_{ac}Q_{bd}g^{cd}-2G_{ac}S_{bd}g^{cd}+h^{ce}h_e^dG_{ac}G_{bd}-2h^{cd}G_{ac}Q_{bd}     \right]t^at^b  +\mathcal{O}(\tau^3) \leq 0
\end{equation}
where
\begin{equation}
    \label{1st term}
   \begin{split}
        &Q_{ac}Q_{bd}g^{cd}t^at^b=\left[ (R_{ac})_L-\frac{1}{2}R_L g_{ac}-\frac{1}{2}R h_{ac}    \right] \left[ (R_{bd})_L-\frac{1}{2}R_L g_{bd}-\frac{1}{2}R h_{bd} \right]g^{cd}t^at^b\\
        &=\Bigg[\frac{1}{4}\Big\{\Big(\nabla_e\nabla_a h^e_c+\nabla_e\nabla_ch_a^e-\square h_{ac}-\nabla_c\nabla_ah  \Big)\left(\nabla_f\nabla_b h^f_d+\nabla_f\nabla_dh_b^f-\square h_{bd}-\nabla_d\nabla_bh  \right)\Big\}g^{cd}\\
        & -\frac{1}{2}\Big\{\Big(\nabla_c\nabla_d h^{cd}-\square h -R_{cd}h^{cd} \Big)\Big(2\nabla_e\nabla_ah_b^e-\square h_{ab}-\nabla_a\nabla_bh  \Big)\Big\}+\frac{R}{2}(\nabla_c\nabla_d h^{cd}-\square h -R_{cd}h^{cd} )h_{ab}\\
        &+\frac{1}{4}(\nabla_c\nabla_d h^{cd}-\square h -R_{cd}h^{cd} )^2 g_{ab}+\frac{1}{4}R^2h_{ac}h_{bd}g^{cd}-Rh_{ac}g^{cd} \Big(\nabla_e\nabla_b h^e_d+\nabla_e\nabla_dh_b^e\\
        &-\square h_{bd}-\nabla_d\nabla_bh  \Big) \Bigg] t^at^b         
        \end{split}
\end{equation}
and 
\begin{equation}
    \label{2nd term}
    \begin{split}
        -2&G_{ac}S_{bd}g^{cd}t^at^b= -2\Bigg[R_b^c\Big[\nabla_d\left\{h_e^d \left(\Gamma^e_{ac} \right)_L\right\} -h_e^d\nabla_a\left( \Gamma^e_{dc} \right)_L -\frac{1}{2}\nabla_dh_{ae}(\nabla^dh_c^e-\nabla^eh^d_c)\\
        &-\frac{1}{2}\{ \nabla_dh(\Gamma^d_{ac})_L+\nabla_ah_e^d(\Gamma^e_{cd})_L\}  \Big]-\frac{1}{2}R_a^d h_{bd}\left(\nabla_c\nabla^eh^c_e -\square h-R_{ec}h^{ec}  \right)\\
        &-\frac{1}{2}R\Big[\nabla_d\left\{h_e^d \left(\Gamma^e_{ab} \right)_L\right\} -h_e^d\nabla_a\left( \Gamma^e_{db} \right)_L -\frac{1}{2}\nabla_dh_{ae}(\nabla^dh_b^e-\nabla^eh^d_b)\\
        &-\frac{1}{2}\Big\{ \nabla_dh(\Gamma^d_{ab})_L+\nabla_ah_e^d(\Gamma^e_{bd})_L\Big\}  \Big]+\frac{1}{4}R h_{ab}\left(\nabla_c\nabla_dh^{cd}-\square h-R_{cd}h^{cd}\right)\\
        &-\frac{1}{2} \left(R_a^d  g_{bd}-\frac{1}{2}R g_{ab}  \right)\mathcal{P}\Bigg]t^at^b
    \end{split}
\end{equation}
with 
\begin{equation}
    \label{P}
    \begin{split}
        \mathcal{P}=\frac{1}{4}\Big[&4R^{ab}h_{ac}h^c_b+4h^{ab}\square h_{ab}+4h^{ab}\nabla_a\nabla_b h-4 h^{ab}\nabla_c\nabla_b h^c_a-4h^{ab}\nabla_a\nabla_ch^c_b+\nabla_a h\nabla_b h^{ab}\\
        &-\nabla_ah \nabla^a h-4\nabla_b h^b_a\nabla_c h^{ac}+3\nabla_a h_{bc}\nabla^a h^{bc}\Bigg]
    \end{split}
\end{equation}
and 
\begin{equation}
    \label{3rd term}
    h^{ce}h_e^d G_{ac}G_{bd}t^at^b=h^{ce}h_e^d\left( R_{ac}-\frac{1}{2}R g_{ac} \right) \left( R_{bd}-\frac{1}{2}R g_{bd} \right)t^at^b~~.
\end{equation}
 Lastly,  \begin{equation}
     \label{4th term}
     -2h^{cd}G_{ac}Q_{bd}t^at^b=-2h^{cd}\left( R_{ac}-\frac{1}{2}R g_{ac} \right) \left[ (R_{bd})_L-\frac{1}{2}R_L g_{bd} -\frac{1}{2}R h_{bd} \right] t^at^b
 \end{equation}

Similarly, the NEC at order $\mathcal{O}(h^2)$ yields 
\begin{equation}
    \label{2nd order NEC}
    \begin{split}
        \Big[&\nabla_d\left\{h_e^d \left(\Gamma^e_{ab} \right)_L\right\} -h_e^d\nabla_a\left( \Gamma^e_{db} \right)_L -\frac{1}{2}\nabla_dh_{ae}(\nabla^dh_b^e-\nabla^eh^d_b)\\
        &-\frac{1}{2}\Big\{ \nabla_dh(\Gamma^d_{ab})_L+\nabla_ah_e^d(\Gamma^e_{bd})_L\Big\}  \Big]n^an^b+\mathcal{O}(\tau^3)\leq 0~.
    \end{split}
\end{equation}
Noting that $R_{ab}, ~ R^a_b,~ R  $ vanish for an asymptotically flat spacetime, the dominant energy condition  given by equation \ref{2nd order DEC} simplifies to 

\begin{equation}
    \label{simple 2nd order DEC}
    \begin{split}
        &\left[ (R_{ac})_L-\frac{1}{2}R_L g_{ac}    \right] \left[ (R_{bd})_L-\frac{1}{2}R_L g_{bd} \right]g^{cd}t^at^b\\
        =&\Bigg[\frac{1}{4}\Big\{\Big(\nabla_e\nabla_a h^e_c+\nabla_e\nabla_ch_a^e-\square h_{ac}-\nabla_c\nabla_ah  \Big)\left(\nabla_f\nabla_b h^f_d+\nabla_f\nabla_dh_b^f-\square h_{bd}-\nabla_d\nabla_bh  \right)\Big\}g^{cd}\\
        & -\frac{1}{2}\Big\{\Big(\nabla_c\nabla_d h^{cd}-\square h  \Big)\Big(2\nabla_e\nabla_ah_b^e-\square h_{ab}-\nabla_a\nabla_bh  \Big)\Big\}+\frac{1}{4}(\nabla_c\nabla_d h^{cd}-\square h )^2 g_{ab} \Bigg] t^at^b \leq 0,        
    \end{split}
\end{equation}
up to order $\mathcal{O}(h^2)$. On the other hand the null energy condition given by equation \ref{2nd order NEC} remains the same up to order $\mathcal{O}(h^2)$. 

It is important to emphasize that the four constraints derived from the energy conditions, Eqs.~\ref{SEC}--\ref{NEC}, are completely general and apply to an arbitrary metric perturbation $h_{ab}$. No assumption regarding the origin of the perturbation is required at this stage. However, upon specializing to perturbations generated by BMS transformations, $h_{ab}=\mathcal{L}_{\eta}g_{ab}$, a distinctive feature emerges: the DEC and NEC constraints are identically satisfied at order $\mathcal{O}(h)$. Consequently, nontrivial restrictions from these conditions arise only at order $\mathcal{O}(h^{2})$, necessitating a second-order analysis to extract additional physical constraints.

\section{BMS Energy Conditions}{\label{sec-bms energy}}

Up to this stage, the formalism has been developed without imposing any specific structure on the metric perturbation $h_{ab}$. In general, such a perturbation may originate from a wide range of geometric deformations, including genuine dynamical distortions of the spacetime, coordinate redefinitions, or large gauge transformations. In the present analysis, however, we restrict our attention to perturbations generated by the action of a diffeomorphic vector field $\eta^{a}$ on a background metric $g_{ab}$. The perturbation is therefore taken to be of the form
\begin{equation}
h_{ab} = \mathcal{L}_{\eta} g_{ab}, 
\qquad 
\bar g_{ab} = g_{ab} + \mathcal{L}_{\eta} g_{ab},
\end{equation}
so that the transformed geometry is obtained through an infinitesimal BMS transformation.

From this point onward, we specialize exclusively to perturbations arising from BMS transformations. Accordingly, the vector field $\eta^{a}$ is understood to belong to the BMS algebra, and the perturbation is fully characterized by the corresponding supertranslation parameter $f$. The constraints derived in the previous sections therefore translate into restrictions on the allowed BMS generators compatible with the chosen background geometry. To obtain explicit and physically meaningful conditions on $f$, it is necessary to specify the metric structure of both $g_{ab}$ and the transformed geometry $\bar g_{ab}$.

Thus, consider the most general metric, expressed in $(u,r,\theta,\phi)$ coordinates in the BMS form, as
\begin{equation}
    \label{gbs}
    \begin{split}
        ds^2\Bigg|_{\bar{g}_{ab}}=&-\left(1-\frac{2m}{r}\right)du^2-dudr+r^2(d\theta^2+\sin^2\theta d\phi^2  )\\
        &+r C_{\theta\theta}d\theta^2+r C_{\phi\phi}d\phi^2+2r C_{\theta\phi}d\theta d\phi^2+2g_{u\theta}dud\theta+2g_{u\phi}dud \phi
    \end{split}
\end{equation}

Among the class of background geometries capable of yielding the form given in Eq.~\ref{gbs}, the simplest and most widely studied example is the Schwarzschild spacetime. We therefore specialize to the case where the background metric $g_{ab}$ is taken to be the Schwarzschild solution. In this setting, the previously derived constraints translate into explicit conditions on the supertranslation parameter $f$. In what follows, we determine the restrictions that must be imposed on $f$ so that the BMS-transformed Schwarzschild geometry described by Eq.~\ref{gbs} remains physically admissible.

Consequently, we consider the Schwarzschild metric, expressed in  $(u,r,\theta,\phi)$ coordinates as

\begin{equation}
    \label{schwarzschild}
    ds^2\Bigg|_{g_{ab}}=-\left(1-\frac{2m}{r}  \right)du^2-2dudr+r^2(d\theta^2+\sin^2\theta d\phi^2)
\end{equation}
We further consider the class of diffeomorphisms that have the large-$r$ falloffs
\begin{equation}
    \label{falloff}
    \eta^u,\eta^r\sim \mathcal{O}(r^0) ~~\text{and}~~\eta^\theta,\eta^\phi\sim \mathcal{O}\left(\frac{1}{r}\right)
\end{equation}
These conditions are necessary for the vector field to be $\mathcal{O}(r^0)$ at large $r$ in an orthonormal frame, thereby eliminating boosts and rotations that grow with $r$ at infinity.

We next consider the class of diffeomorphisms that maintain the asymptotic structure of the
Schwarzschild metric \ref{schwarzschild}, and satisfies the following set of conditions

\begin{equation}
    \label{lie condition}
    \begin{split}
        \mathcal{L}_\eta g_{rr}&=0\\
        \mathcal{L}_\eta g_{rA}&=0\\
        \mathcal{L}_\eta g_{ur}&=0\\
        \gamma^{AB}\mathcal{L}_\eta g_{AB}&=0
    \end{split}
\end{equation}
The vector field satisfying the above gauge conditions \ref{lie condition} and the large-$r$ falloff behavior \ref{falloff} is given by
\begin{equation}
    \label{eta}
    \eta=f\partial_u+\frac{1}{2}D^2 f-\frac{1}{r}\left[(\partial_\theta f)\partial_\theta+\left(\frac{\partial_\phi f}{\sin^2\theta}  \right)\partial_\phi   \right]
\end{equation}
with 
\begin{equation}
    \label{laplacian}
    D^2\mathcal{F}(\theta,\phi)=\frac{1}{\sin\theta}\partial_\theta\left( \sin\theta\partial_\theta\mathcal{F} \right)+\frac{1}{\sin^2\theta}\partial_\phi^2\mathcal{F}
\end{equation}
representing the Laplacian of any function $\mathcal{F}(\theta,\phi)$ on the $2$-sphere and $f=f(\theta,\phi)$ parametrizing the transformation in equation \ref{eta}, known as supertranslation parameter.

With the action of this vector field $\eta$ on the remaining metric components, we have the transformed Schwarzschild metric as 
\begin{equation}
    \label{gbs sch}
    \begin{split}
        &ds^2\Bigg|_{\bar{g}_{ab}}=(g_{ab}+\mathcal{L}_{\eta}g_{ab})dx^adx^b=-\left(1-\frac{2m}{r} \right)du^2-2dudr+r^2(d\theta^2+\sin^2\theta d\phi^2)\\
        &+r(D^2f-2\partial_\theta^2f)d\theta^2+r(\sin^2\theta D^2 f-2\sin\theta\cos\theta\partial_\theta f-2\partial_\phi^2f)d\phi^2-2r\partial_\theta\partial_\phi\left(f+\frac{f}{\sin^2\theta}\right)d\theta d\phi\\
    &+2\partial_\theta\left[\left(\frac{2m}{r}-1 \right)f-\frac{1}{2}D^2f  \right]dud\theta+2\partial_\phi\left[\left(\frac{2m}{r} -1 \right)f-\frac{1}{2}D^2f  \right]dud\theta
    \end{split}
\end{equation}

Comparison of equation \ref{gbs sch} with equation \ref{gbs} gives
\begin{equation}
    \label{coefficients}
    \begin{split}
        C_{\theta\theta}&=D^2f-2\partial_\theta^2 f\\
        C_{\phi\phi}&=\sin^2\theta D^2 f-2\sin\theta\cos\theta\partial_\theta f-2\partial_\phi^2f\\
        C_{\theta\phi}&=-\partial_\theta\partial_\phi\left(f+\frac{f}{\sin^2\theta}\right)\\
        g_{u\theta}&=\partial_\theta\left[\left(\frac{2m}{r}-1 \right)f-\frac{1}{2}D^2f  \right]\\
        g_{u\phi}&=\partial_\phi\left[\left(\frac{2m}{r}-1 \right)f-\frac{1}{2}D^2f  \right]\\
    \end{split}
\end{equation}

giving the metric perturbation $h_{ab}$ as

\begin{equation}
    \label{hab}
h_{ab}=\begin{bmatrix}
0 & 0 & g_{u\theta} & g_{u\phi}  \\
0 & 0 & 0 & 0  \\
g_{u\theta} & 0 & r C_{\theta\theta} & r C_{\theta\phi}  \\
g_{u\phi} & 0 & r C_{\theta\phi} & r C_{\phi\phi} \\
\end{bmatrix}
\end{equation}
Note that Eqs.~\ref{coefficients} and \ref{hab} follow from the particular large-$r$ falloff and Lie-derivative conditions displayed in Eqs.~\ref{falloff} and \ref{lie condition}. Different choices of asymptotic behaviour or gauge conditions will alter the explicit form of these components, and the procedure for deriving constraints on the generator $\eta^{a}$ via the energy conditions will remain unchanged.

To evaluate the energy conditions in the geometry defined by Eqs.~\ref{gbs sch} and \ref{hab} we introduce a timelike four-velocity $t^{a}$ and a null four-vector $n^{a}$. For a stationary, comoving observer we take
\[
t^{a}=\bigl(1/\sqrt{1-2m/r},\,0,\,0,\,0\bigr),
\]
so that, using $\bar g_{ab}$, the covariant components become
\[
t_{a}=\bigl(-\sqrt{1-2m/r},\,-1/\sqrt{1-2m/r},\,g_{u\theta}/\sqrt{1-2m/r},\,g_{u\phi}/\sqrt{1-2m/r}\bigr),
\]
which satisfies $t^{a}t_{a}=-1$. For a radial null trajectory we choose
\[
n^{a}=(0,\,1,\,0,\,0),\qquad n_{a}=(-1,\,0,\,0,\,0),
\]
so that $n^{a}n_{a}=0$. These choices are convenient for projecting the transformed Ricci tensor and Einstein tensor onto timelike and null congruences when checking the SEC, WEC, DEC and NEC in the BMS-transformed Schwarzschild geometry.

Thus, with $R_{ab}=0$ and $R=0$, we have from equation \ref{SEC},
\begin{equation}
\label{general SEC}
     (2 m-r) \left\{C_{\theta \theta} +\csc ^2\theta ~ C_{\phi \phi }\right\}+2 r \left\{\partial_\theta g_{u\theta }+\cot\theta ~ g_{u\theta }+\csc ^2\theta \partial_\phi g_{u\phi}\right\}\geq 0~,
\end{equation}
 which with equations \ref{coefficients} and \ref{hab} gives
\begin{equation}
    \label{final SEC}
    \begin{split}
        &2 \csc ^2\theta \partial_\phi^2 f \left(-2 m+2 r \csc ^2\theta -r\right)+r \Bigg[-\partial_\theta^2 f+\partial_\theta^4 f+2 \cot \theta \partial_\theta^3 f+\csc ^4\theta   \partial_\phi^4 f\\
        &+\csc ^2\theta  \Big\{-\partial_\theta^2  f+2 \partial_\theta^2\partial_\phi^2f+\cot \theta  \left(\partial_\theta f-2 \partial_\theta \partial_\phi^2 f\right)\Big\}\Bigg]\leq 4 m \left(\partial_\theta^2 f+\cot \theta  \partial_\theta f\right)~,
    \end{split}
\end{equation}
for the strong energy condition.  

Since we are interested in the asymptotic form in the  limit $r\to \infty$, equation \ref{final SEC} can be written as 

\begin{equation}
    \label{limit final SEC}
    \begin{split}
        2 \csc ^2\theta (2 \csc ^2\theta-1)\partial_\phi^2 f&+\Bigg[-\partial_\theta^2 f+\partial_\theta^4 f+2 \cot \theta \partial_\theta^3 f+\csc ^4\theta   \partial_\phi^4 f\\
        &+\csc ^2\theta  \Big\{-\partial_\theta^2  f+2 \partial_\theta^2\partial_\phi^2f+\cot \theta  \left(\partial_\theta f-2 \partial_\theta \partial_\phi^2 f\right)\Big\}\Bigg]\leq 0~.
    \end{split}
\end{equation}
Similarly, from equation \ref{WEC}, we have for the weak energy condition,
\begin{equation}
\label{general WEC}
   \begin{split}
        &m (2 m-r) \left\{C_{\theta \theta }+\csc ^2\theta C_{\phi \phi }\right\}+2 m \left\{\partial_\theta g_{u\theta }+\cot \theta  g_{u\theta }+\csc ^2\theta  \partial_\phi g_{u\phi }\right\}\\
        &\geq r (r-2 m) \Bigg[\csc ^2\theta  \Big\{\partial_\phi^2 C_{\theta \theta }-2 \left(\partial_\theta \partial_\phi C_{\theta \phi }+\cot \theta \partial_\theta C_{\phi \phi }\right)+\partial_\theta^2 C_{\phi \phi }\\
        &+\left(2 \csc ^2\theta -1\right) C_{\phi \phi}+2 \partial_\phi g_{u\phi}\Big\}-\cot \theta \partial_\theta C_{\theta \theta }+C_{\theta \theta }+2 \partial_\theta g_{u\theta }+2 \cot \theta  g_{u\theta }\Bigg]~,
   \end{split}
\end{equation}
which with the substitutions from equations \ref{coefficients} and \ref{hab} gives
\begin{equation}
    \label{final WEC}
    \begin{split}
        &2 \cot \theta  \left[\left\{m^2 (4 r+2)-r^3\right\} \partial_\theta f-m r \partial_\theta^3 f\right]+2 \csc ^2\theta  \partial_\phi^2f \Bigg[(2 m+r) (2 m r+m-r^2)\\
        &-2 r \csc ^2\theta  \left\{3 r \csc ^2\theta  (r-2 m)+6 m r+m-3 r^2\right\}\Bigg]+r \csc ^2\theta  \Big[ m \partial_\theta^2 f-2 \left(2 m r+m-r^2\right) \partial_\theta^2\partial_\phi^2 f\Big]\\
        &+8 r^2 \cot \theta  \csc ^4\theta  (r-2 m) \partial_\theta\partial_\phi^2f+r \cot \theta  \csc ^2\theta  \left[2 \left(4 m r+m-2 r^2\right) \partial_\theta\partial_\phi^2 f-m \partial_\theta f\right]\\
        &+m \{m (8 r+4)+r\}\partial_\theta^2 f+r \csc ^4\theta  \left[-m \partial_\phi^4 f-2 r (r-2 m) \partial_\phi^2\partial_\theta^2f\right]\geq m r \partial_\theta^4 f+2 r^3 \partial_\theta^2 f~.
    \end{split}
\end{equation}
Thus, in the asymptotic limit $r\to \infty$, equation \ref{final WEC}  gives the constraint,
\begin{equation}
    \label{limit final WEC}
    \begin{split}
        \partial_\theta^2f+\left(\csc ^2\theta +6 \cot ^2\theta  \csc ^4\theta \right)\partial_\phi^2 f+\cot \theta  \Big[\partial_\theta f+\csc ^2\theta  \left\{\cot \theta \partial_\theta^2\partial_\phi^2 f+\left(2-4 \csc ^2\theta \right)\partial_\theta\partial_\phi^2 f\right\}\Big]\geq 0
    \end{split}
\end{equation}
originating from the WEC.

For the specific choice of the Schwarzschild background, the dominant and null energy conditions are identically satisfied at linear order in the perturbation $\mathcal{O}(h)$. In other words, the $\mathcal{O}(h)$ contributions to the DEC and NEC vanish for the class of BMS transformations considered here, so that Eqs.~\ref{DEC} and \ref{NEC} are trivially fulfilled at this order.

The first nontrivial contributions from DEC and NEC therefore appear at next-to-next-to-leading order (NNLO) in $h_{ab}$. Specializing to the Schwarzschild metric \eqref{schwarzschild} with the perturbation $h_{ab}$ given in \eqref{hab} and imposing the Lie conditions \eqref{lie condition}, the DEC at NNLO reduces, in the asymptotic $r\to\infty$ limit, to the constraint 
\begin{equation}
   \label{psi_zeta NNLO DEC}
   \begin{split}
        \psi(\theta,\phi)\zeta(\theta,\phi)\geq 0
   \end{split}
\end{equation}
with
\begin{equation}
    \label{psi}
    \begin{split}
        \psi(\theta,\phi)&=2 \partial_\phi^2 f+\partial_\theta^2  f-2\partial_\theta^2\partial_\phi^2 f+\sin 2 \theta  ~\partial_\theta F-\cos 2 \theta ~ \partial_\theta^2 f+4 \cot \theta ~ \partial_\theta\partial_\phi^2f\\
        &+12 \csc ^4\theta ~\partial_\phi^2 f-12 \csc ^2\theta ~\partial_\phi^2 f+2 \csc ^2\theta~\partial_\phi^2\partial_\theta^2 f-8 \cot \theta \csc ^2\theta \partial_\phi^2\partial_\theta f 
    \end{split}
\end{equation}
and 
\begin{equation}
    \label{zeta}
    \begin{split}
       \zeta(\theta,\phi)= \partial_ \theta^2 f&+\cot \theta \partial_ \theta f+\csc ^2\theta \partial_ \phi^2 f+6 \cot ^2 \theta  \csc ^4\theta  \partial_\phi^2 f+\cot ^2\theta  \csc ^2\theta \partial_ \theta^2\partial_ \phi^2 f\\
        &-4 \cot \theta  \csc ^4\theta \partial_ \theta\partial_ \phi^2 f+2 \cot \theta  \csc ^2\theta \partial_ \theta\partial_ \phi^2 f
    \end{split}
\end{equation}
and, from equation \ref{2nd order NEC}, the NEC at NNLO reduces to
\begin{equation}
  \label{NNLO NEC}
  \begin{split}
        \Big[-\partial_\theta^2 f+\cot \theta ~\partial_\theta f+\csc ^2\theta ~\partial_\phi^2 f\Big]^2+\frac{1}{16} \csc ^8\theta  &\Big[(\sin 3 \theta -7 \sin \theta )\partial_\theta\partial_\phi f\\&+8 \cos \theta\partial_\phi  f\Big]^2\geq 0
  \end{split}
\end{equation}

It is important to emphasize that, in deriving Eq.~\ref{NNLO NEC} corresponding to the NEC, all $r$-dependent terms cancel exactly. As a consequence, the resulting constraint is purely angular and does not rely on the limit $r\to\infty$. The NEC therefore yields the strongest constraint on the supertranslation parameter $f(\theta,\phi)$ within the class of BMS transformations considered here.

For the Schwarzschild background, where \(R_{ab}=R=0\), the dominant energy condition is identically satisfied at linear order   $\mathcal{O}(h)$ in the perturbation. This outcome is physically natural as the DEC enforces non-negative energy density and a causal energy flux, so its preservation under generic BMS transformations indicates that several fundamental properties of the Schwarzschild geometry survive the transformation. In particular, causal energy flow is maintained (ruling out superluminal energy propagation) and the DEC is consistent with the non-negativity of the ADM energy \cite{schoen1979proof, schoen1981proof, witten1981new}.

Similarly, the NEC is preserved at next-to-leading order \(\mathcal{O}(h)\) ensuring that the effective Hamiltonian is bounded from below. The fact that both DEC and NEC hold through \(\mathcal{O}(h)\) reinforces the requirement that any BMS-generated perturbation \(\bar g_{ab}=g_{ab}+h_{ab}\) should also satisfy the SEC and WEC (Eqs.~\ref{final SEC}, \ref{final WEC} or, more generally, Eqs.~\ref{SEC}, \ref{WEC}). 
The particular choice of series expanding $\bar{R}_{ab}$ and $\bar{R}$ in terms of $R_{ab}$, $R$ and $h_{ab}$ enables us to demonstrate that some of the intrinsic properties of the Schwarzschild black hole carry on to the newly transformed metric,  evident from the meeting of DEC and NEC up to NLO.
Although our expansion can be carried to arbitrarily high order, in the asymptotic regime \(|g_{ab}|\gg|h_{ab}|\) higher-order corrections are parametrically suppressed. Accordingly, we expect that the leading- and next-to-leading-order constraints capture the significant constraints on the admissible BMS generators.

\section{Discussion and Conclusion}{\label{disc}}
The infinite-dimensional character of the BMS group has long been regarded as one of the most intriguing features of asymptotically flat spacetimes. Yet an immediate conceptual question arises. Does mathematical admissibility automatically imply physical realizability? In other words, should every formally allowed supertranslation be regarded as physically acceptable, or must additional criteria intervene?

In this work, we have developed a toolkit to identify and remove unphysical metric perturbations, including those generated by BMS transformations, that are prohibited by the classical energy conditions. Rather than treating supertranslations as unrestricted gauge redundancies, we impose physically motivated constraints rooted in the requirement that gravity remains attractive, energy densities remain non-negative, local energy flow
vector to be non-spacelike and causality to be preserved. When these conditions are enforced on BMS-transformed geometries, a nontrivial restriction emerges on the otherwise arbitrary supertranslation function space. The resulting picture clearly indicates that asymptotic symmetry, while being infinite-dimensional at the purely geometric level, is substantially reduced once physical admissibility is demanded.

An important and general conceptual feature of our analysis is to impose the energy conditions in a perturbative framework in a generally transformed metric space, $g_{ab}\to g_{ab}+h_{ab}$.  In doing so, we have expanded the curvature tensors and the Ricci scalar in terms of the metric perturbation, which have been used  in the rest of the work. In particular, our analysis provides a comprehensive study of the corresponding series expansions, which can be used in different fields of gravity and cosmology, including the study of gravitational waves. This perspective allows one to carry over the intrinsic geometric properties of the background spacetime while systematically analyzing how perturbation modifies physical viability. The procedure is quite general, applicable to any metric undergoing perturbation, including BMS transformations with $h_{ab}=\mathcal{L}_{\eta}g_{ab}$ .

Furthermore, to exemplify and to obtain the explicit form of the energy conditions, we considered the perturbed metric as the most general geometry expressed in $(u,r,\theta,\phi)$ coordinate system in the standard BMS form, with the background given by the Schwarzschild metric. As a result, the supertranslation function that parametrizes the transformation acquires constraints since its angular dependence undergoes restrictions by the requirement that the strong and weak energy conditions remain satisfied up to order $\mathcal{O}(h)$. Interestingly, the null and dominant energy conditions  are identically  preserved at order $\mathcal{O}(h)$, indicating that not all energy conditions are equally sensitive to supertranslation freedom. Nontrivial restrictions emerge only beyond the leading order, revealing a hierarchy in how physical constraints act on asymptotic symmetries.

This leads to a broader interpretational point in that the traditional BMS construction identifies an infinite-dimensional symmetry algebra at null infinity. Our results suggest that the physically admissible subset of this algebra may be significantly smaller once energy conditions are imposed. The symmetry group remains infinite-dimensional in a formal sense, but its physically allowed parameter space is substantially constrained. Thus, the apparent arbitrariness of the supertranslation sector is moderated by classical gravitational consistency requirements.

More generally, the approach adopted here provides a bridge between asymptotic symmetry analysis and classical energy condition arguments. It demonstrates that tools originally designed to constrain matter models and gravitational collapse can also be used to refine our understanding of large gauge transformations. This perspective may have implications for conserved charges, memory effects, and the physical interpretation of soft degrees of freedom, particularly in contexts where asymptotic symmetries play a central role.

In summary, while the BMS group is mathematically vast, physical viability imposes meaningful structure upon it. By demanding that energy conditions remain satisfied after a supertranslation, one restricts the space of permissible transformations and reduces the associated ambiguity. The resulting symmetry space, though still infinite-dimensional, is no longer unrestricted. This suggests that the interplay between asymptotic symmetry and classical gravitational constraints deserves closer attention and may provide further insight into the physical content of large gauge transformations in general relativity.

\section{Acknowledgement} 
Nihar Ranjan Ghosh is supported through a Research Fellowship from the Ministry of
Human Resource Development (MHRD), Government of India.

%\bibliographystyle{unsrt}
%\bibliographystyle{IEEEtran}
%\bibliography{my6}

\end{document}